\documentclass[twocolumn,prb]{revtex4}
\usepackage{graphicx}    
\usepackage{dcolumn}     
\usepackage{bm}          
\newcommand{\Vec}[1]{\mbox{\boldmath$#1$}}
\begin{document}
\title{High $T_c$ superconductivity due to coexisting wide and narrow bands:\\
A fluctuation exchange study on the Hubbard ladder as a test case}
\author{
Kazuhiko Kuroki$^1$, Takafumi Higashida$^1$, 
and Ryotaro Arita$^{2}$\cite{AR}
}
\affiliation{
$^1$ Department of Applied Physics and
Chemistry, The University of Electro-Communications,
Chofu, Tokyo 182-8585, Japan\\
$^2$ Department of Physics, 
University of Tokyo, Hongo 7-3-1, Tokyo 113-0033, Japan}
\date{\today}
\begin{abstract}
We propose that 
when the Fermi level lies within a wide band 
and also lies close to but not within a coexisting narrow band, 
high $T_c$ superconductivity may take place due to 
the large number of interband pair scattering channels and 
the small renormalization of the quasiparticles.
We show using fluctuation exchange method that this mechanism 
works for the Hubbard model on a ladder lattice with 
diagonal hoppings. 
From this viewpoint, we give a possible explanation for the low 
$T_c$ for the actual hole doped ladder compound, 
and further predict a higher $T_c$ for the case of electron doping.
\end{abstract}
\pacs{PACS numbers: }
\maketitle

The discovery of high $T_c$ superconductivity in the 
cuprates\cite{BM}, followed by discoveries of various unconventional 
superconductors, has brought up renewed fascination for the search of 
high $T_c$ superconductors, for which a theoretical guiding principle 
is highly desired. One way to attack this problem is to theoretically 
search for lattice structures that provide good conditions for Cooper pairing.
Ladder-like structures may be considered as a candidate for this,
\cite{DR}, but up to now, $T_c$ in the actual hole doped ladder 
compound Sr$_{14-x}$Ca$_x$Cu$_{24}$O$_{41}$ (14-24-41) 
remains to be around $\simeq 12$ K,\cite{Uehara} which is 
low compared to the layered cuprates.
In this context, 
we have previously proposed a high $T_c$ mechanism due to 
disconnected Fermi surfaces, which can in fact be realized with a 
ladder-like lattice structure with a larger hopping integral in the 
rung direction than in the leg direction,\cite{KA2} 
but up to now there exist  
no actual materials (or methods) to realize the situation we have proposed.
In the present study, we propose a different mechanism
for high $T_c$ superconductivity in systems with coexisting 
narrow and wide bands, which may be realized in actual 
ladder compounds with {\it electron} doping.

\begin{figure}[htb]
\begin{center}
\scalebox{0.6}{
\includegraphics[width=10cm,clip]{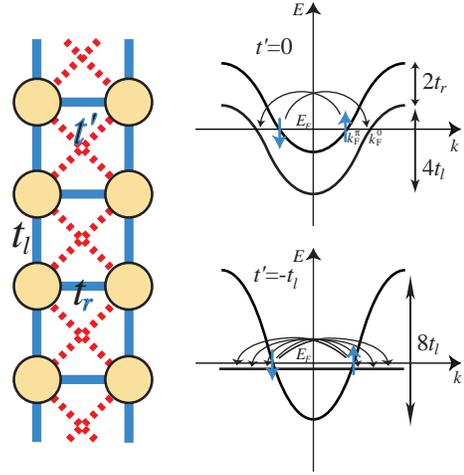}}
\caption{
Left panel: The hopping integrals in the ladder lattice is shown.
Right panel : the energy dispersion for the ladder lattice with 
$t'=0$ (top) and with $t'=-t_l$ (bottom). 
The curved arrows show the pair scattering processes that give rise to 
superconductivity. \label{fig1}}
\end{center}
\end{figure}
\begin{figure}[htb]
\begin{center}
\scalebox{0.6}{
\includegraphics[width=10cm,clip]{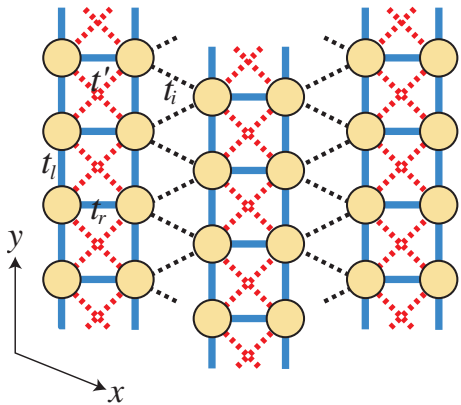}}
\caption{The trellis lattice adopted in the present study. 
\label{fig2}}
\end{center}
\end{figure}
\begin{figure}[htb]
\begin{center}
\scalebox{0.6}{
\includegraphics[width=10cm,clip]{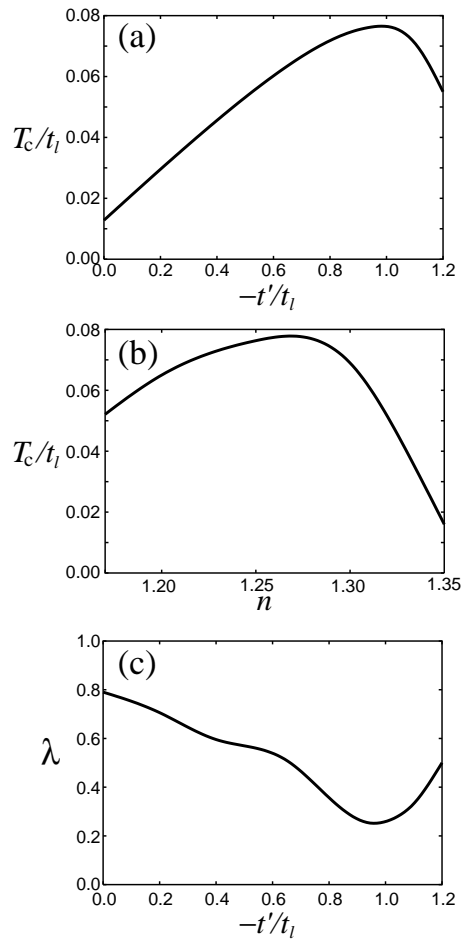}}
\caption{$T_c$ plotted (a) as a function of $-t'$ for $n=1.25$, 
and (b) as a function of $n$ for $-t'=0.95t_l$. (c)The eigenvalue of the 
Eliashberg equation $\lambda$ plotted as a function of $-t'$
for $n=0.9$ and $T=0.05t_l$.\label{fig3}}
\end{center}
\end{figure}
\begin{figure}[htb]
\begin{center}
\scalebox{0.7}{
\includegraphics[width=10cm,clip]{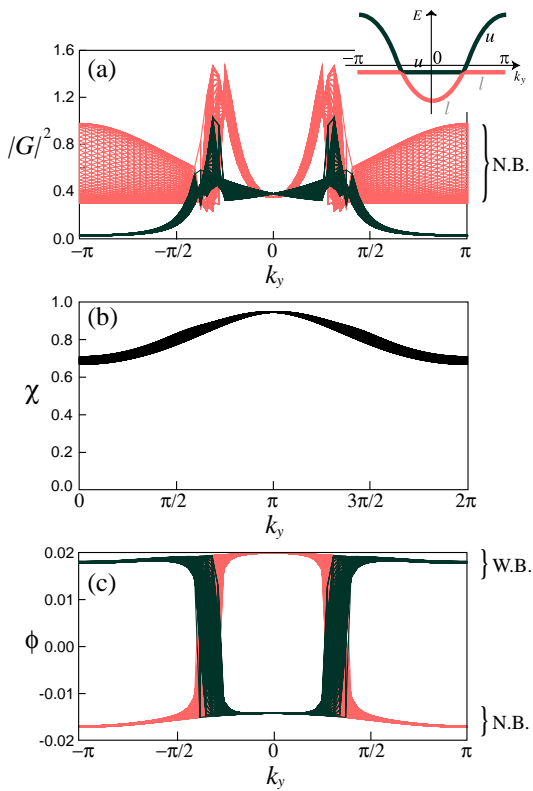}}
\caption{
Color online. (a) $|G_\alpha(k_x,k_y,i\pi k_BT)|^2$, (b) $\chi(k_x,k_y,0)$, 
(c) $\phi_\alpha(k_x,k_y,i\pi k_BT)$ for $n=1.25$, $t'=-0.95t_l$, 
and $T=0.08t_l$  viewed from the direction of the $k_x$ axis.
The thickness of the curves represents the dispersion in the $k_x$ direction.
In (a) and (c), quantities 
for $\alpha=u$ (upper band) and $l$ (lower band) are shown by 
dark green and red curves, respectively, while N.B (W.B.) denotes the 
portions corresponding to those for the narrow band B (wide band A).
The relation between the narrow/wide bands and the lower/upper bands is  
shown in the inset of figure (a).
\label{fig4}}
\end{center}
\end{figure}
\begin{figure}[htb]
\begin{center}
\scalebox{0.6}{
\includegraphics[width=10cm,clip]{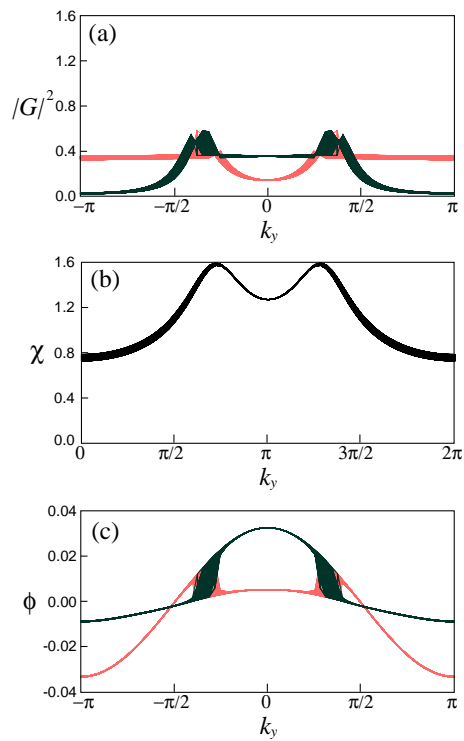}}
\caption{
Plots similar to Fig.\protect\ref{fig4} for 
$n=0.9$ and $t'=-0.95t_l$.\label{fig5}}
\end{center}
\end{figure}
\begin{figure}[htb]
\begin{center}
\scalebox{0.6}{
\includegraphics[width=10cm,clip]{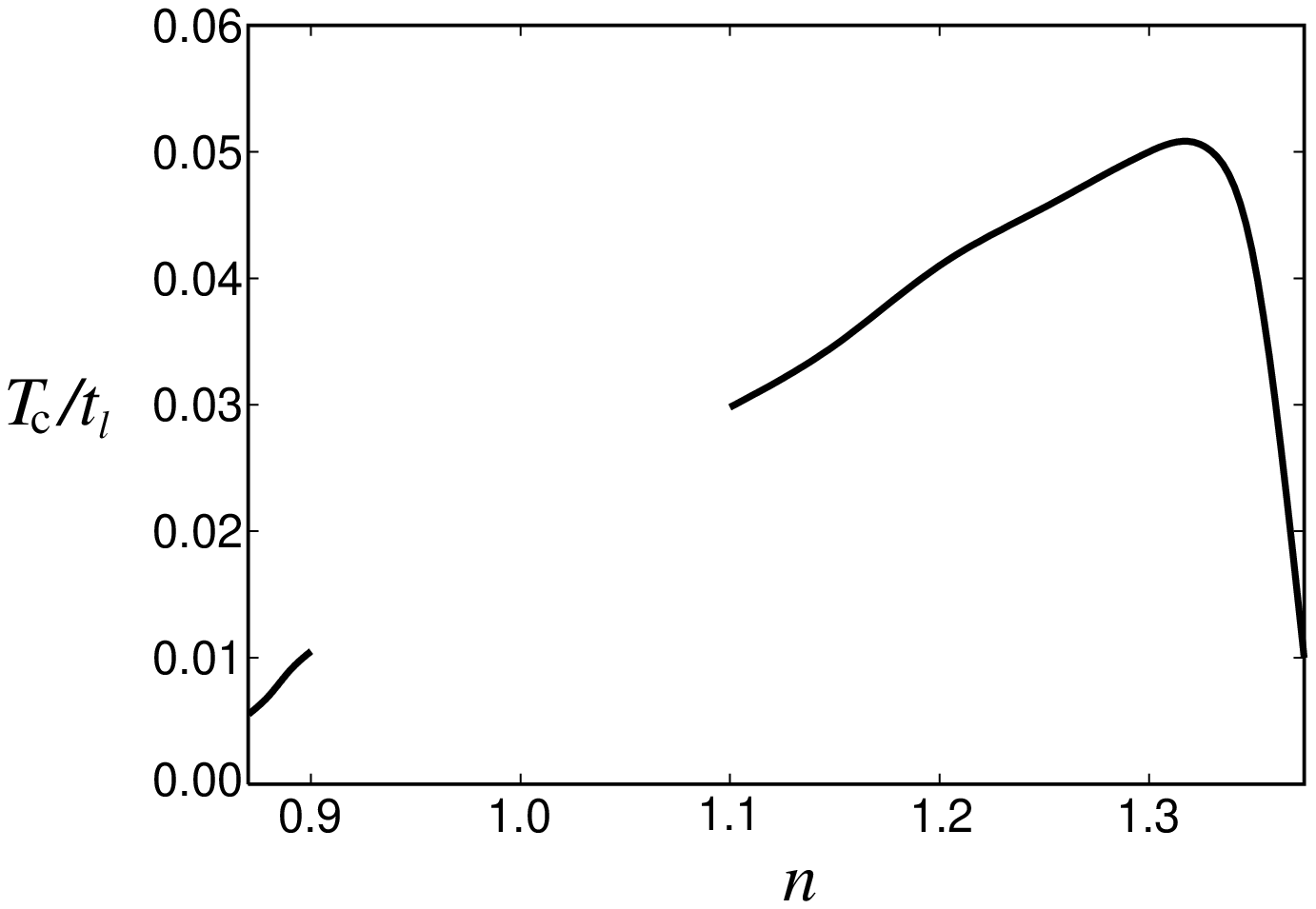}}
\caption{
$T_c$ plotted as a function of $n$ for $t'=-0.4t_l$.\label{fig6}}
\end{center}
\end{figure}

Our idea is as follows. 
Let us consider a system where the Fermi level $E_F$ lies within a band with 
a moderate width (we call this ``the wide band A'') and also 
lies close to, but not within, a narrow 
band (``band B''). If the amplitude of the 
pair scattering processes from band A to band B is strong enough,
the pairing instability may become very large because (i) the 
sign change of the gap, necessary for repulsive pairing interactions, 
occurs between the bands, so that there are no gap nodes on the Fermi surface,
and (ii) there is a huge 
number of interband pair scattering channels due to the narrow character of 
band B (see the bottom of the 
right panel in Fig.\ref{fig1}). Another point to be 
stressed here is that the Fermi level is not {\it within} the narrow 
band, so that the renormalization of the quasiparticles at the Fermi level 
is not so large as to strongly suppress superconductivity. 
Systems with coexisting wide and narrow bands may be reminiscent of models 
consisting of  wide $s$ or $p$ bands and 
narrow $d$ or $f$ bands, but in that case, the amplitude of the 
interband (interorbital) 
pair scattering processes may not be so large because of the 
different character of the orbitals.
Here we consider systems with {\it one orbital} per site,  
where the multiplicity of the bands originates 
from the lattice structure rather than the multiplicity of 
the orbitals at each site, so that the pairing interaction can  
be strong, originating from the large on-site repulsion.

The above condition for the energy bands 
can most simply be satisfied in a 
tightbinding model on a ladder lattice 
with diagonal hoppings shown in Fig.\ref{fig1}.
The Hamiltonian of this model is given in momentum space as 
\[
H_{\rm kin}=\sum_{k\sigma} 
(c_{k\sigma}^\dagger, d_{k\sigma}^\dagger)
\left(
\begin{array}{cc}
-2t_l\cos k & -2t'\cos k-t_r\\
-2t'\cos k-t_r & -2t_l\cos k
\end{array}
\right)
\left(
\begin{array}{c}
c_{k\sigma}\\
d_{k\sigma}
\end{array}
\right)
\]
where $c_{k\sigma}$ and $d_{k\sigma}$ 
annihilate an electron with 
spin  $\sigma$ at wave number $k$ on the left and right legs, respectively, 
and $t_l$, $t_r$, and $t'$ are the hopping integrals in the leg, rung, and 
diagonal directions, respectively. The dispersion of the two bands are given 
as 
$
\varepsilon_{\pm}(k)=-2(t_l\pm t')\cos k\mp t_r.
$
When $t'=0$ (Fig.\ref{fig1}, top of the right panel), 
the two bands have identical dispersions with a level offset of 
$2t_r$, while one of the bands (band B) is  narrower
than the other (band A) in the presence of $t'$, and becomes 
perfectly flat for $t'=\pm t_l$ (Fig.\ref{fig1}, bottom of the right panel).

Here, we consider the on-site interaction ($U$) term 
in addition to the above kinetic energy terms, 
and also take into account the trellis-like 
lattice structure of the actual cuprate ladder compounds,
where the ladders are weakly coupled by diagonal hoppings $t_i$
\cite{Kontani,commentti} 
We estimate the superconducting transition temperature
of this Hubbard model using  
the combination of the fluctuation exchange method (FLEX) and 
the Eliashberg equation, which has been successfully applied to the 
problem of layered high $T_c$ cuprates.\cite{Bickers,Dahm,Grabowski} 
The values of $U$, $t_r$, $t_i$ will be fixed at $U=6t_l$, $t_r=t_l$, 
and  $t_i=0.25t_l$ throughout the study.
We define the the band filling $n$ as 
$n=$[number of electrons]/[number of sites], so when the bands are both
fully filled, the band filling is $n=2$.

In the two band version of FLEX,\cite{Kontani} the Green's function $G$, 
the susceptibility $\chi$, the self-energy $\Sigma$, and 
the superconducting gap function $\phi$ all become $2\times 2$ matrices,
e.g., $G_{lm}(\Vec{k}, i\varepsilon_n)$, where $l,m$ specify
the two sites in a unit cell.
The orbital-indexed matrices for Green's function and the gap functions
can be converted into band-indexed 
ones with a unitary transformation. 
As for the spin susceptibility, we diagonalize the 
spin susceptibility matrix and 
concentrate on the larger eigenvalue, denoted as $\chi$.  

The actual calculation proceeds as (i) Dyson's equation is 
solved to obtain the renormalized Green's function $G(k)$,
where $k\equiv(\Vec{k},i\epsilon_n)$ denotes the 2D wave-vectors and 
the Matsubara frequencies,
(ii) the effective electron-electron interaction $V^{(1)}(q)$ 
is calculated by collecting RPA-type bubbles and ladder diagrams consisting
of the renormalized Green's function, namely, 
by summing up powers of the irreducible susceptibility 
$\chi_{\rm irr}(q)\equiv -\frac{1}{N}\sum_k G(k+q)G(k)$ 
($N$:number of $k$-point meshes),
(iii) the self energy is obtained as 
$\Sigma(k)\equiv\frac{1}{N}\sum_{q} G(k-q)V^{(1)}(q)$, 
which is substituted into Dyson's equation in (i), 
and these procedures are repeated until convergence is attained.

We determine $T_c$ as the temperature at which the eigenvalue $\lambda$ of 
the linearized Eliashberg equation, 
\[
\lambda\phi_{l m}(k)\nonumber = -\frac{T}{N}\sum_{k'l'm'}
\times V_{l m}^{(2)}(k-k')G_{ll'}(k')G_{mm'}(-k')\phi_{l'm'}(k'),
\]
reaches unity.
Here the pairing interaction $V^{(2)}$ for singlet pairing 
is given by $V^{(2)}(q)=
U+\frac{3}{2}U^2\chi_{\rm irr}(q)/(1-U\chi_{\rm irr}(q))
-\frac{1}{2}U^2\chi_{\rm irr}(q)/(1+U\chi_{\rm irr}(q))$.

Throughout the study, we take up to $64\times 64$ $k$-point meshes and 
the Matsubara frequencies $\epsilon_n$ from 
$-(2N_c-1)\pi T$ to $(2N_c-1)\pi T$ with $N_c$ up to 8192 in order to
ensure convergence at low temperatures.

We now move on to the results.
We first show the $t'$ dependence of $T_c$ for $n=1.25$. We consider 
the case of $t'<0$ since this is the realistic choice of sign for the 
cuprates.\cite{KAA} 
In Fig.\ref{fig3}(a), we plot $T_c$ as a function of $-t'$ for 
$n=1.25$. It can be seen that $T_c$ takes its maximum around $-t'=t_l$, 
where band B is flat. 
There, $T_c$ almost reaches $0.08t_l$, which is extremely high if $t_l$ is 
assumed to be of the order of few hundred meV as in the cuprates.

To trace back the origin of this high $T_c$, 
we look into the Green's functions, spin susceptibility and the 
gap functions for $-t'=0.95t_l$ and $n=1.25$.
In Fig.\ref{fig4}, we plot 
$|G_{u,l}(k_x,k_y,i\pi T)|^2$, $\chi(k_x,k_y,0)$, and 
$\phi_{u,l}(k_x,k_y,i\pi T)$ viewed from the direction of the 
$k_x$ axis. Here, we display the results for $G$ and $\phi$ 
in the band representation, 
where $u$ and $l$ denote the upper and 
lower portions of the bands  
as shown in the inset of Fig.\ref{fig4}(a).
Since bands A and B intersect with each other, 
the wide band A (the narrow band B) consists of 
the upper (lower) band around $k_y=\pm\pi$ and the lower (upper) 
band around $k_y=0$.
In the Green's functions, 
the two peak-like structures seen around $k_y=\pm\pi/4$ 
is due to the Fermi level crossing, 
while the narrow structure (noted as N.B.) owes to the flatness of band B.
Since the volume of the Fermi surface (the length $2k_F$) is $\simeq \pi/2$, 
the wide band A is nearly quarter filled, and thus 
the narrow band is fully filled, $i.e.$, the Fermi level 
lies above the narrow band. 
The spin susceptibility has a broad structure again 
due to the flatness of band B, which enhances the number of pair scattering 
channels due to spin fluctuations. The gap function changes sign between 
band A and band B, but there is no sign change within each band,
as expected in our intuitive picture.

In Fig.\ref{fig3}(b), we fix $-t'$ at $0.95t_l$ and show the 
band filling dependence of $T_c$, 
which takes its maximum around 
$n=1.25$. This result 
shows that $T_c$ becomes low when $n$ is too large, that is,  
when the Fermi level lies too far 
above the narrow band since this will make the system close to a 
purely single band model.  $T_c$ also goes down when the Fermi level 
comes too close to or within the narrow band B.

In fact, when the Fermi level is {\it within}
the narrow band, we find a $t'$ dependence that is completely the opposite 
to what is seen in Fig.\ref{fig3}(a).
In Fig.\ref{fig3}(c), we plot the eigenvalue $\lambda$
of the Eliashberg equation for $n=0.9$ at $T=0.05t_l$. In this case, a finite 
$T_c$ is not obtained for all of the $t'$ values, 
so we plot $\lambda$ for a fixed temperature 
instead of $T_c$.
As seen in the figure, $\lambda$ takes its {\it minimum} around 
$-t'=t_l$. In this case, the Fermi level is within band B
as can be seen from the Green's function shown in Fig.\ref{fig5}(a).
Namely, the length between the two peaks is about 
$0.7\pi$ (which is in fact larger than the case of $n=1.25$ meaning that 
the band is not rigid), so that both of the bands 
has to be partially occupied.\cite{commentfilling}
We can see in Fig.\ref{fig5}(a) that the Green's functions are  
small (the quasiparticle renormalization is strong) compared to 
those in Fig.\ref{fig4}(a), so this should suppress $T_c$.
Moreover, the gap function (Fig.\ref{fig5}(c)) changes 
sign within the wide and the narrow bands near the Fermi surface, 
and this also should work destructive against superconductivity. 
The sign change in the gap means that contributions from the 
pair scattering interactions 
{\it within} the narrow (and the wide) band are large,  
which is a consequence of the  Fermi level crossing of the 
narrow band with a large density of states. This means that the 
peak structure in the susceptibility 
(Fig.\ref{fig5}(b)) originates 
from both interband and intraband scattering processes.

The above results suggest that superconductivity with high 
$T_c$ may be obtained in the ladder compounds 
for the case of {\it electron} doping.\cite{holecomment}
The condition for the maximum $T_c$, $-t'\simeq t_l$, is of 
course unrealistic for the cuprate ladders, but we notice in Fig.\ref{fig3}
that the enhancement of $T_c$ remains even if we deviate 
from $-t'=t_l$, and relatively high $T_c(\sim 0.04t_l)$ 
is obtained even below $-t'=0.5t_l$. 
Note that 
$t'$ should be around $t'\simeq -0.4t_l$ in the 
actual cuprate ladder compounds, 
assuming that $t'$ has values similar to 
those for the layered cuprates such as YBa$_2$Cu$_3$O$_7$\cite{KAA}. 

To look in more detail into the possibility of 
high $T_c$ superconductivity for 
realistic values of the hopping integrals, 
we plot in Fig.\ref{fig6} the band filling 
dependence of $T_c$ for $-t'=0.4t_l$.
$T_c$ is not calculated near $n=1$ because (i) antiferromagnetic 
fluctuations strongly develop at high temperatures so that $T_c$ 
is not obtained, and in any case, 
(ii) FLEX loses its validity in the vicinity of half filling,
where a Mott transition should take place.
We find that a maximum $T_c$ of $\sim 0.05t_l$, 
which is still considerably 
high, is reached at around $n=1.3$, namely, when a large amount of 
electrons is doped. By contrast, for hole doping $(n<1)$, 
$T_c$ turns out to be much smaller,
namely, of the order $0.001t_l$, which is 
of the same order as the $T_c(\simeq$12K) for the 
actual 14-24-41 compound.\cite{Uehara}
The origin of the difference 
between the electron doped and the hole doped cases,
can again be traced back to the Green's functions, 
the spin susceptibility,  and the gap functions,
in which the characteristic features seen for 
$t'=-0.95t_l$ (Fig.\ref{fig4}) still remain to some extent. The 
details on this point will be published elsewhere.\cite{KHA2}


We have seen that our high $T_c$ mechanism due to 
coexisting narrow and wide bands works for a ladder system.
Now, an interesting question is to ask how general this mechanism is.
We have in fact found similar results 
for the Hubbard model on a lattice where a pair of square lattices 
with in-plain nearest neighbor hoppings ($t$) is coupled by out-of-plain 
vertical ($t_v$) 
and diagonal ($t'$) hoppings.\cite{KHA2} 
The Hamiltonian of this model is given as 
These results seem to suggest that the present mechanism is likely 
to work on two band lattices having the 
Hamiltonian of the form 
\[
H_{\rm kin}=\sum_{\Vec{k},\sigma} 
(c_{\Vec{k},\sigma}^\dagger, d_{\Vec{k},\sigma}^\dagger)
\left(
\begin{array}{cc}
\varepsilon(\Vec{k}) & \alpha\varepsilon(\Vec{k})+\beta\\
\alpha\varepsilon(\Vec{k})+\beta & \varepsilon(\Vec{k})
\end{array}
\right)
\left(
\begin{array}{c}
c_{\Vec{k},\sigma}\\
d_{\Vec{k},\sigma}
\end{array}
\right),
\]
where $\alpha$ and $\beta$ are constants. 
Here, a flat band coexists with a wide band for $\alpha=\pm 1$
($\varepsilon(k)=-2t_l\cos k$ for the ladder, and 
$-2t(\cos k_x+\cos k_y)$ for the coupled planes).
It would be an interesting future problem to investigate the 
validity of the present superconducting mechanism in a more wide class of 
models where wide and narrow bands coexist.

To summarize, we have 
proposed a mechanism for high $T_c$ superconductivity in 
a two band system where wide and narrow bands coexist. 
When the Fermi level lies close to but not within the narrow band, 
$T_c$ becomes higher as the flatness of the narrow band increases,
while completely the opposite takes place when the Fermi level is 
within the narrow band.
From this viewpoint, we have given a possible explanation for the 
rather low $T_c$ for the actual hole doped 14-24-41 ladder compound, 
and have further predicted a higher $T_c$ for the case of electron doping.
As for the 14-24-41 compound, recent NMR experiments have observed 
a coherence peak and an unchanged Knight shift across $T_c$.\cite{Fujiwara} 
As for the presence of the coherence peak, 
the singlet gap function obtained in our study 
can be consistent with the experiments since the gap does not change its 
sign {\it on} the Fermi surface. 
As for the unchanged Knight shift, 
we believe further investigation is necessary, but even this is due to 
spin-triplet pairing originating from effects not included in the present study
(e.g., phonons), we have to explain why $T_c$ of the singlet pairing
is even lower than 12K.
The present viewpoint can still be relevant for 
answering this question, and we believe 
that the prediction for a higher $T_c$ in the 
case of electron doping remains valid.

KK acknowledges H. Aoki and J. Akimitsu for 
valuable discussion.
Numerical calculation has been performed
at the facilities of the Supercomputer Center,
Institute for Solid State Physics,
University of Tokyo.

%


\begin{thebibliography}{99}
\bibitem[*]{AR} Present address: 
Max Planck Institute for Solid State Research, 
Heisenbergstr. 1, Stuttgart 70569, Germany
\bibitem{BM} J.G. Bednorz and K.A. M\"{u}ller, Z. Phys. B, Condens. Matter
{\bf 64} 189 (1986).
\bibitem{DR} E. Dagotto and T.M. Rice, Science {\bf 271}, 618 (1996)  
and references therein.
\bibitem{Uehara} M. Uehara, T. Nagata, J. Akimitsu, H. Takahashi, 
N. M\^{o}ri, and K. Kinoshita, J. Phys. Soc. Jpn. {\bf 65}, 2764 (1996).
\bibitem{KA2} K. Kuroki and R. Arita, Phys. Rev. B {\bf 64}, 024501 (2001).
\bibitem{Kontani} H. Kontani and K. Ueda, Phys. Rev. Lett. 
{\bf 80}, 5619 (1998).
\bibitem{commentti} The introduction of $t_i$ makes the narrow band slightly 
dispersive even for $t'=\pm t_l$.
\bibitem{Bickers} N.E. Bickers,  D.J. Scalapino, and S.R. White, 
Phys. Rev. Lett. {\bf 62}, 961 (1989).
\bibitem{Dahm} T. Dahm and L. Tewordt, Phys. Rev. B {\bf 52}, 1297 (1995).
\bibitem{Grabowski} S. Grabowski,  J.Schmalian, M. Langer, and K.H. Bennemann, 
Solid State Commun. {\bf 98}, 611 (1996).
\bibitem{KAA} See, e.g., K. Kuroki, R. Arita, H. Aoki, 
Phys. Rev. B {\bf 60}, 9850 (1999), and references therein.
\bibitem{commentfilling}
Since $n<1$, neither of the bands can be fully filled.
On the other hand, if there were completely no electrons in one of the 
bands, the volume of the Fermi surface of 
the other would be $0.9\pi$ for $n=0.9$.
\bibitem{holecomment} The Fermi level can be brought {\it below} 
and close to the narrow band in the hole doped case, but for $t'<0$,
this occurs when holes are significantly doped away from half filling 
so that the band filling is very small. 
In such a case, $T_c$ enhancement due to 
electron correlation effects cannot be expected.
\bibitem{KHA2} K. Kuroki and R. Arita, unpublished. 
\bibitem{Fujiwara} M. Fujiwara, N. M\^{o}ri, Y. Uwatoko, T. Matsumoto, 
N. Motoyama, and S. Uchida, Phys. Rev. Lett. {\bf 90}, 137001 (2003).
\end{thebibliography}
\end{document}